\documentclass[aps,prb,twocolumn,showpacs,preprintnumbers,amsmath,amssymb,superscriptaddress]{revtex4-1}

\usepackage{graphicx}
\usepackage{dcolumn}
\usepackage{bm}
\usepackage{color}
\usepackage{ulem}

\newcommand{\expect}[1]{\langle #1 \rangle}

\newcommand{\labs} {\left\vert}
\newcommand{\rabs} {\right\vert}

\newcommand{\lab} {\left\langle}
\newcommand{\rab} {\right\rangle}

\newcommand{\be}{\begin{equation}}
\newcommand{\ee}{\end{equation}}
\newcommand{\bea}{\begin{eqnarray}}
\newcommand{\eea}{\end{eqnarray}}

\newcommand{\tSO}{t_{\rm SO }}

\begin{document}

\title{Edge magnetism impact on electrical conductance and thermoelectric properties of graphenelike nanoribbons}

\author{Stefan Krompiewski}
\email{stefan@ifmpan.poznan.pl} \affiliation{Institute of Molecular Physics, Polish Academy of
Sciences, 60-179 Pozna\'n, Poland}

\author{Gianaurelio Cuniberti}
\affiliation{ Institute for Materials Science and Max Bergmann Center of Biomaterials,
TU Dresden, 01062 Dresden, Germany}
\affiliation{Center for Advancing Electronics Dresden, TU Dresden, 01062 Dresden,
Germany}
\affiliation{Dresden Center for Computational Materials Science (DCMS), TU Dresden, 01062 Dresden, Germany}

\date{\today}

\begin{abstract}

Edge states in narrow quasi two-dimensional nanostructures determine, to a large extent, their electric, thermoelectric and magnetic properties. Non-magnetic edge states may quite often lead to topological insulator type behavior. However another scenario develops when the zigzag edges are magnetic and the time reversal symmetry is broken. In this work we report on the electronic band structure modifications, electrical conductance and thermoelectric properties of narrow zigzag nanoribbons with spontaneously magnetized edges. Theoretical studies based on the Kane-Mele-Hubbard tight-binding model show that for silicene, germanene and stanene both the Seebeck coefficient and the thermoelectric power factor are strongly enhanced for energies close to the charge neutrality point. Perpendicular gate voltage lifts the spin degeneracy of energy bands in the ground state with antiparallel magnetized zigzag edges and makes the electrical conductance significantly spin-polarized. Simultaneously the gate voltage worsens the thermoelectric performance. Estimated room-temperature figures of merit for the aforementioned nanoribbons can exceed a value of 3 if phonon thermal conductances are adequately reduced.
\end{abstract}

\pacs{73.63.Kv,72.25.-b,73.50.Lw}



\maketitle

\section{Introduction}

Recently there has been enormous interest in the physical properties of graphene nanostructures and other similar quasi two-dimensional systems. It is believed that these systems will soon enter the field of modern nanoelectronics, including spintronics and calorytronics. Here we study the latter two issues.
The question how to improve the thermoelectric performance of nanostructures has been addressed in many scientific reports. It is worth mentioning in this context the following  methods aimed at achieving this purpose: shape and grain boundary manipulations [\onlinecite{Sevi12,TUDPRB15}], energy spectrum engineering [\onlinecite{Karamitaheri12,Zheng15}], and magnetism-related concepts based on magnetic proximity effects coming from substrates (staggered magnetization) and the impact of the external magnetic field [\onlinecite{Wierzb15}]. The existence of edge magnetism in the case of narrow high quality graphene nonoribbons is now well documented [\onlinecite{Joly10,Tao11,Gao13,Magda14,Li14}]. It is very probable that similar evidences in support of edge magnetism in other graphenelike nanostructures will also be demonstrated soon. The best known graphenelike nanostructures (e.g. silicene, germanene and stanene) are quasi two-dimensional rather than strictly 2-dimensional because their sublattices are shifted with respect to each other by the so-called buckling distance in the off-plane direction [\onlinecite{Guzman2007,Cahangirov2009,Liu2011,Kamal2015,Ezawa2015,Gomez16}]. In contrast to graphene the buckled structures usually have a non-negligible intrinsic spin-orbit coupling which strongly influences their electronic energy band structures and physical properties of interest here.


\section{Model and Methodology}

In order to catch the essential  physics of graphenelike nanoribbons (GLNRs) we use a tight-binding Hamiltonian with
Hubbard-type electron correlations, intrinsic spin-orbit interaction (ISOI), and an extra term describing the effect of the gate voltage applied in a perpendicular way. Noteworthily, the presence of the ISOI introduces anisotropy, so it is necessary to include in the Hubbard part of the Hamiltonian both the spin diagonal correlations and the off-diagonal ones. The total Hamiltonian reads:

\begin{eqnarray} \label{H0}
  H_0 &=&  - \!\!\!\! \sum \limits_{ <i j>,\sigma } \!\!\! t_{ij} c_{i\sigma}^\dagger c_{j\sigma}
  \! +H_{SO}\nonumber \\
 &&+ \delta_{\theta,0}H_U^{out}+\delta_{\theta,\pi/2}H_U^{in}+H_V,
\end{eqnarray}
with
\begin{eqnarray} \label{Hso}
H_{SO} &=&  i \, \tSO \!\!\!\! \sum \limits_{ <\!\!<ij>\!\!>} \!\!\!
 \nu_{ij} (c_{i\uparrow}^\dagger c_{j\uparrow}-c_{i\downarrow}^\dagger c_{j\downarrow} ),
\end{eqnarray}
\begin{eqnarray} \label{inout}
H_U^{out} &=& U \sum \limits_{i} (\expect{n_{i
\downarrow}} n_{i\uparrow} + \expect{n_{i \uparrow}} n_{i
\downarrow} - \expect{n_{i \uparrow}} \lab n_{i\downarrow}\rab), \nonumber
\\
H_U^{in} &=& -U \sum \limits_{i}(\expect{S_i^+}
S_i^- + \expect{ S_i^-} S_i^ + - \expect{S_i^+} \expect{ S_i^-}).
\end{eqnarray}

\be \label{HV}
H_V=V \sum \limits_{i,\sigma}\mu_i c_{i\sigma}^\dagger c_{j\sigma} .
\ee

In Eq. (2) the summation runs over the next nearest-neighbors, and the factor $\nu_{ij}=\pm 1$,
depending on whether the path from i to j is clockwise or counterclockwise. The factor $\mu_i$, in turn, equals 1 (-1) for sublattice A (B). Other symbols have the following meaning: $t$ is the hopping integral, $U$ is the intraatomic Coulomb repulsion, $\theta= \pi/2 \;(0)$ corresponds to the in-plane (out-of-plane) configuration,  $c_{i \sigma} \; \; (c_{i \sigma}^+)$ are annihilation (creation) operators, and $i$ and $\sigma$ stand for a lattice site and spin, respectively, whereas
$n_{i \sigma} = c_{i \sigma}^+ c_{i \sigma}$ and $S_i^+ = c_{i \uparrow}^+ c_{i \downarrow}$, $S_i^- = (S_i^+)^\dagger $.
Angular brackets stand for expectation values over the ground state, and

\begin{figure}[b!]
 \centering \includegraphics [width=1.3\columnwidth, trim=5cm 7cm 0cm 4cm,
clip=true] {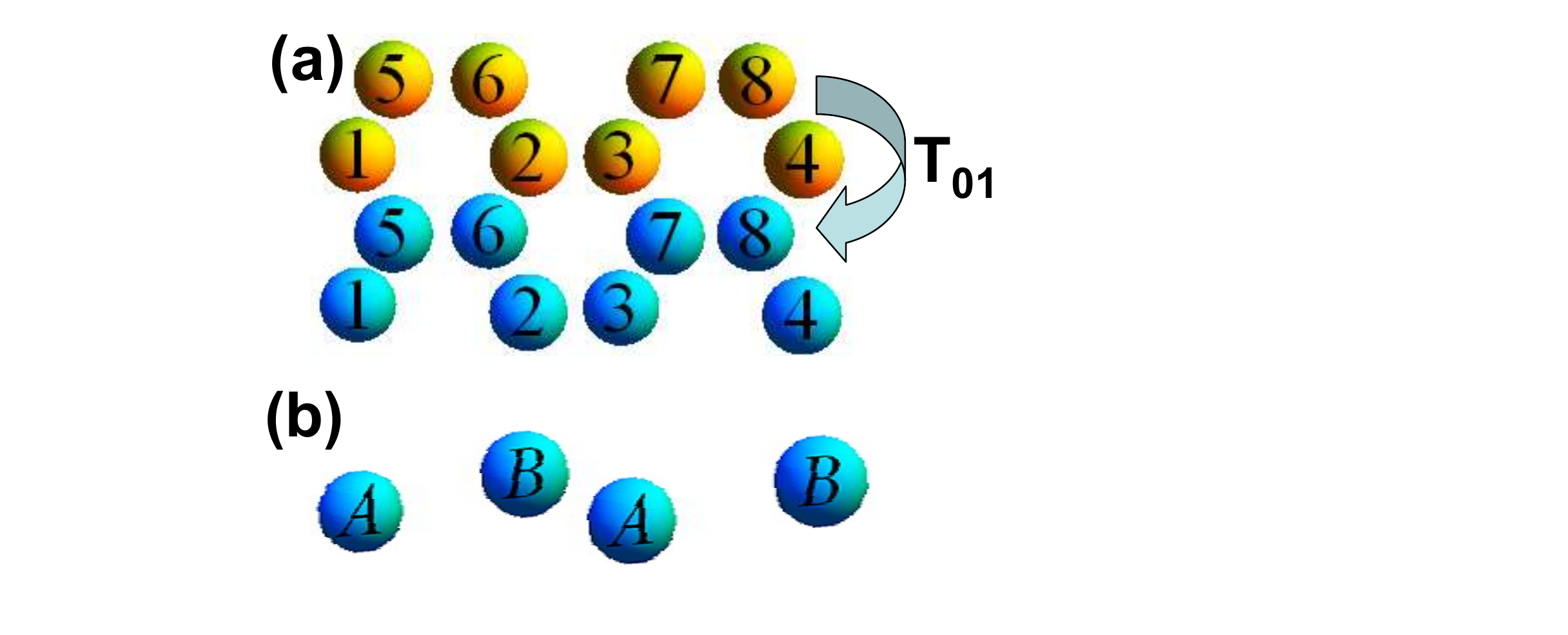}
  \caption{\label{fig1}
  Two neighboring super-cells of a graphenelike nanoribbon with the coupling matrix $T_{01}$ (a), and the side view which shows the buckling of the sublattices (b). The nanoribbon is infinitely long in the zigzag direction.}
\end{figure}

\begin{eqnarray} \label{out}
\expect{n_{i \uparrow}} &=& \frac{1}{2 \pi}\int \limits_{- \pi}^{\pi} u_i^*(E_k)u_i(E_k) f(E_k-\mu) dk, \nonumber
\\
\expect{n_{i \downarrow}} &=& \frac{1}{2 \pi}\int \limits_{- \pi}^{\pi} u_{i+N}^*(E_k)u_{i+N}(E_k)f(E_k-\mu) dk,
\end{eqnarray}

\begin{eqnarray} \label{Spm}
\expect{S_i^+} &=& \frac{1}{2 \pi}\int \limits_{- \pi}^{\pi} u_i^*(E_k)u_{i+N}(E_k)f(E_k-\mu) dk, \nonumber \\
\expect{S_i^-} &=& \frac{1}{2 \pi}\int \limits_{- \pi}^{\pi} u_{i+N}^*(E_k)u_i(E_k)f(E_k-\mu) dk.
\end{eqnarray}

\be \label{out}
m_i^{out} =\mu_B \expect{n_{i \uparrow}-n_{i \downarrow}},
\; \; \;
m_i^{in} =\mu_B \expect{S_i^+ + S_i^-}.
\ee

The electronic band structure is determined from the eigenequation

\be \label{eig}
(H_0+e^{ika} T_{01}+e^{-ika} T_{01}^\dagger)u_k = E_k u_k,
\ee

where the $T_{01}$ matrix describes coupling between 2 neighboring super-cells (as shown in Fig.\ref{fig1}).
Explicitly for 8 atoms in the super-cell:

\begin{eqnarray} \label{T01}
T_{01} = \left(
     \begin{matrix}
      \hat{A}(\zeta) & \hat{0} \\
      \hat{0} & \hat{A}(\bar \zeta)
     \end{matrix}
\right), \nonumber \\
\hat{A}(\zeta) = \left(
     \begin{matrix}
     \zeta &0 &0 &0 &\bar{t} &\bar{\zeta} &0 &0 \\
     0 &\bar{\zeta} &0 &0 &\zeta &\bar{t} &\zeta &0 \\
     0 &0 &\zeta &0 &0 &\bar{\zeta} &\bar{t} &\bar{\zeta} \\
     0 &0 &0 &\bar{\zeta} &0 &0 &\zeta &\bar{t} \\
     0 &0 &0 &0 &\bar{\zeta} &0 &0 &0 \\
     0 &0 &0 &0 &0 &\zeta &0 &0 \\
     0 &0 &0 &0 &0 &0 &\bar{\zeta} &0 \\
     0 &0 &0 &0 &0 &0 &0 &\zeta
     \end{matrix}
\right),
\end{eqnarray}

with $\zeta=i t_{SO}$, $\bar{\zeta}=-\zeta$, and $\bar{t}=-t$ (nearest neighbor hopping).

In order to determine which of the possible configurations constitutes the ground state, grand canonical potentials are computed from the following expression (with a correction due to the last terms in Eqs. \ref{inout})

\be \label{Omega}
\Omega= - \frac{k_B T}{2 \pi} \int \limits_{-\pi}^{\pi} \ln \left( 1+\exp \frac{\mu-E_k}{k_B T}  \right)dk + \rm correction.
\ee

The configurations in question are those with in-plane (IN) or out-of-plane (OUT) magnetization arrangements, and parallel (P) or antiparallel (AP), respectively, magnetic orientations of the opposite zigzag edges (abbreviations: In-AP, Out-AP, IN-P and OUT-P).

\section{Electronic band structure and edge magnetic moments}

\begin{figure}[t!]
\centering \includegraphics [width=1\columnwidth, trim=6cm 1.1cm 7cm 1cm,
clip=true] {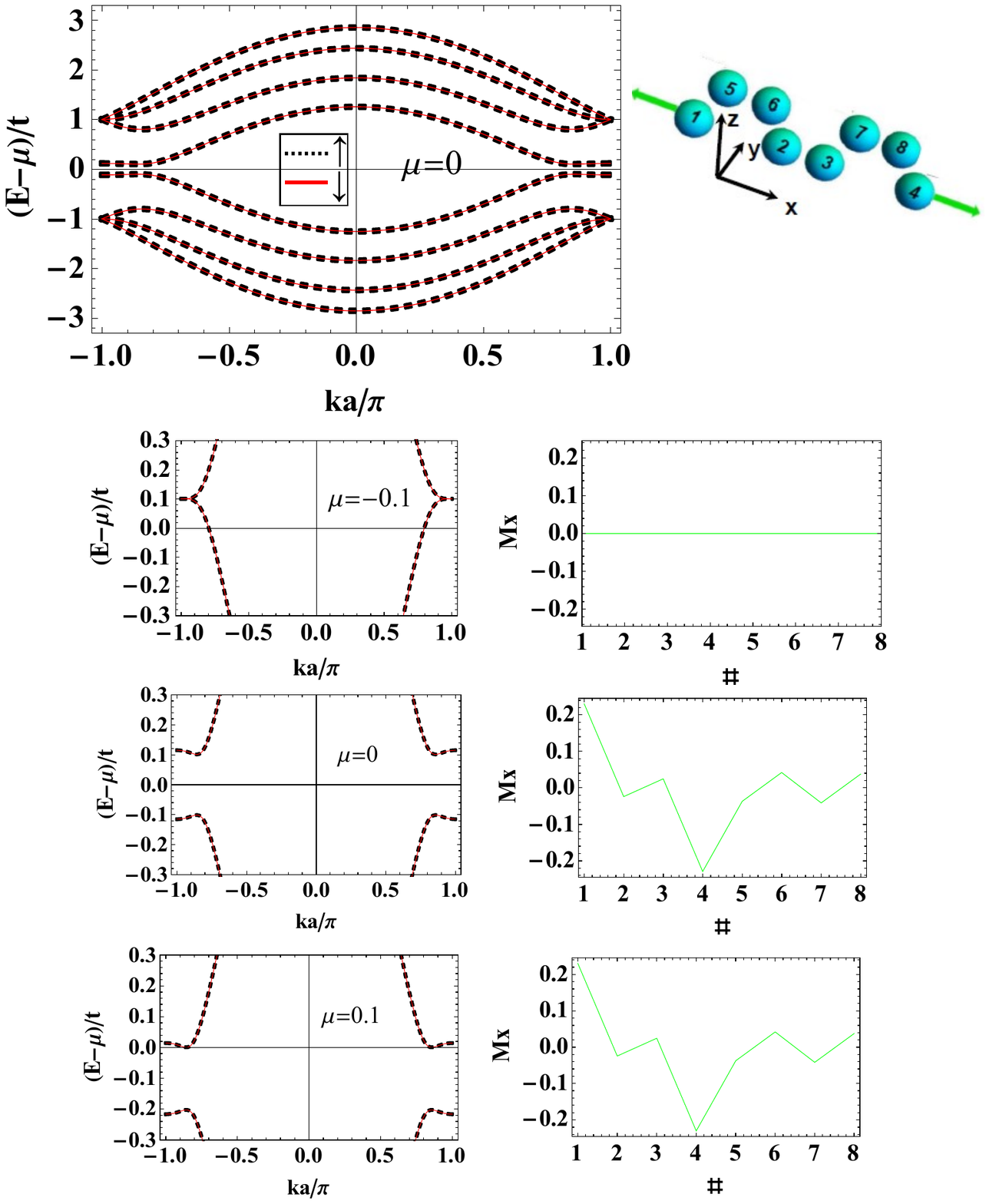}
  \caption{\label{FigPihM}
  Band structures of the IN-AP (in-plane antiparallel) configuration, and the corresponding magnetization profiles (right column) of the narrow stanene nanoribbon. For $\mu$ (in t-units) close to the charge neutrality point the edge atoms 1 and 4 have significant magnetic momments (in $\mu_B$).}
\end{figure}

\begin{figure}[t!]
\centering \includegraphics [width=1\columnwidth, trim=6cm 1.1cm 7cm 1cm,
clip=true] {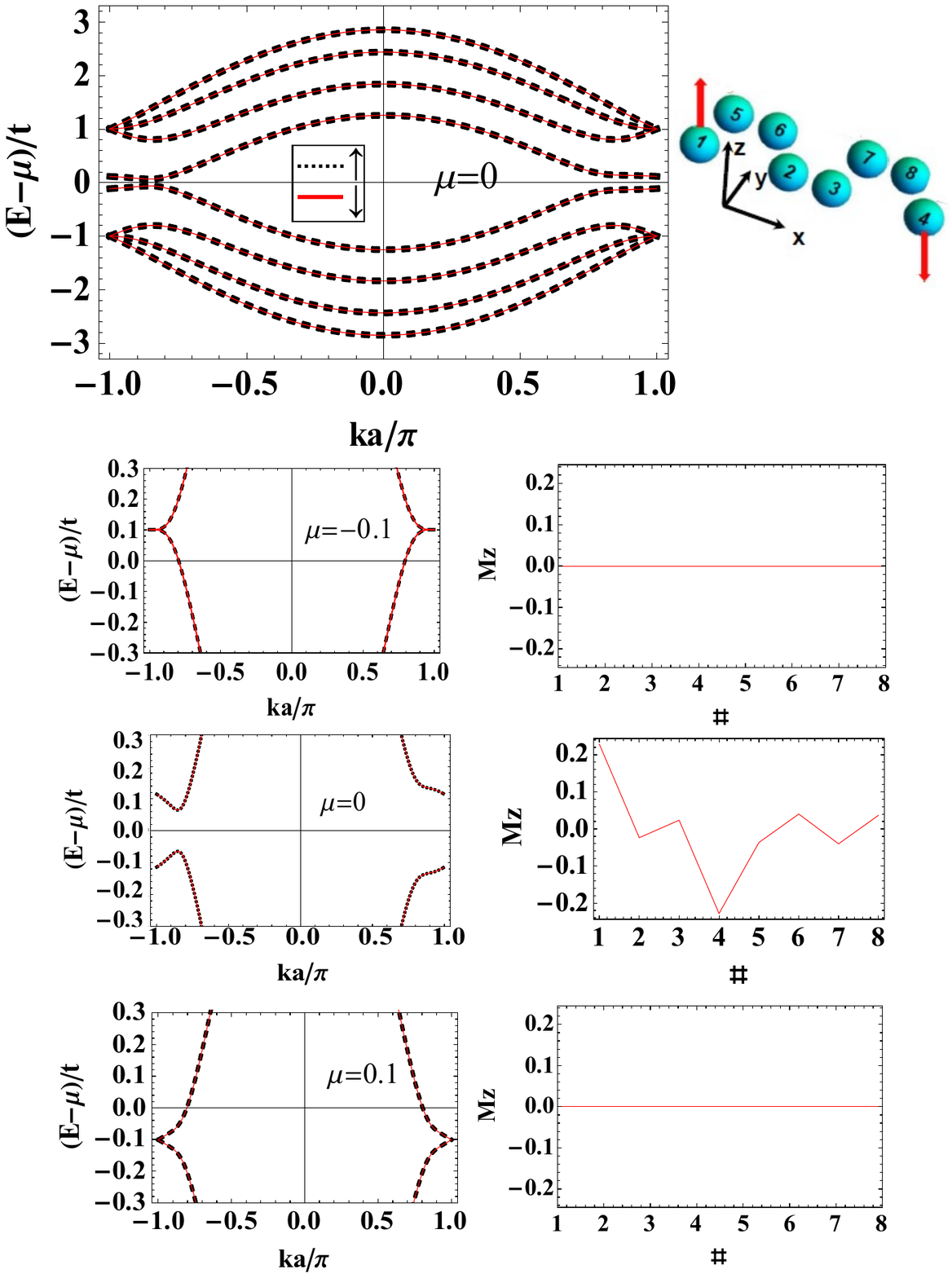}
  \caption{\label{Fig0M}
  As Fig.\ref{FigPihM} but for the Out-AP (out-of-plane antiparallel) configuration.}
\end{figure}

Studies of nanoribbons with magnetic moments are computationally much more demanding than those corresponding to nonmagnetic ones [\onlinecite{SK11}].
 This is so because magnetic moments, and thereby also electronic band structures, depend strongly on the chemical potential position. Hence, for each $\mu$ the band structure has to be determined again and again. This is visualized in Figs \ref{FigPihM} and \ref{Fig0M}, which show that bands in the vicinity of the K and K' points differ from one another depending on whether magnetic moments do or do not exist. Moreover it is readily seen that on the one hand in the Out-AP case the energy spectrum may be  valley-polarized (different energy gaps at K and K' points) and on the other hand the edge magnetism disappears with increasing $\mu$ earlier in that case than in the In-AP configuration (\textit{cf} magnetization profiles for $\mu=0.1$). Incidentally, in the case of graphene there is neither ISOI nor magnetic anisotropy, that is why the configurations IN and OUT are equivalent.

Theoretical studies of graphenelike nanoribbons are usually carried out for the out-of-plane configuration [\onlinecite{Soriano10,Lu14}], but it is now known that the in-plane configuration is often energetically the most stable one [\onlinecite{Lado2014,WBK2015,SKNanotech2016}].  Here most of our attention is directed to this very configuration which constitutes the ground state for energies close to the charge neutrality point (CNP).


\section{Electronic transport and thermoelectric performance}
The following standard relations are used for the electrical conductance, Seebeck coefficient and electronic heat conductance (Ref. [\onlinecite{Karamitaheri12}] and references therein):
\begin{eqnarray} \label{Spm}
G &=& \frac{2 e^2}{h} I_0 \; \; [1/\Omega], \nonumber \\
S &=& - \frac{k_B}{\labs e \rabs} \frac{I_1}{I_0} \; \; [V/K], \nonumber \\
\kappa_{el} &=& \frac{2 T k_B^2}{h} \left(I_2-\frac{I_1^2}{I_0} \right) \; \; [W/K], \nonumber \\
I_j &=& \int \limits_{- \infty}^{\infty}
\left( \frac{E-\mu}{k_B T} \right)^j T_{el}(E) \left(- \frac{\partial f(E-\mu)}{\partial E} \right) dE.
\end{eqnarray}
Above $e, \; h, \; k_B$ and $ \mu_B $ are fundamental physical constants, and T denotes the temperature. G, S, $\kappa_{el}$ stand for the conductance, Seebeck coefficient and electronic thermal conductance (with corresponding units in brackets), respectively. $T_{el}=T_{el,\uparrow}+T_{el,\downarrow}$ is the (ballistic) transmission matrix equal to the number of forward propagating modes. Finally, $f$ is the Fermi-Dirac distribution function, and $\mu$ denotes the chemical potential. The temperature is set equal to 300K, in view of the fact that room temperature magnetic order on zigzag edges of narrow nanoribbons has been recently reported [\onlinecite{Magda14}].

Other quantities of interest here are thermoelectric power factor and figure of merit
$TPF=S^2 G \; [pW/K^2]$ and $ZT=S^2 G T/(\kappa_{el}+\kappa_{ph})$, respectively.

\begin{figure}[h!]
 \centering \includegraphics [width=1\columnwidth, trim=0cm 0cm 2cm 0.5cm,
clip=true] {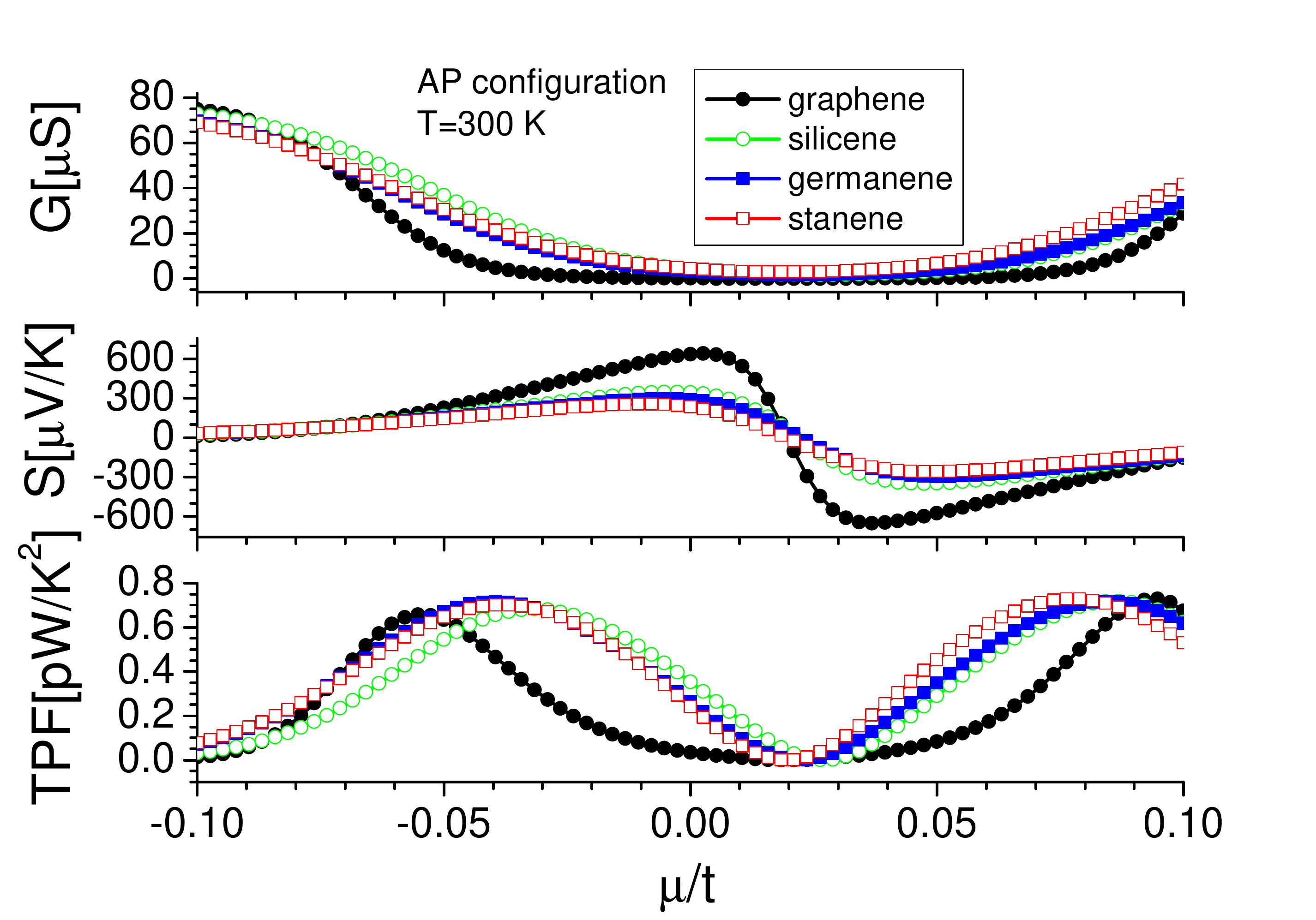}
  \caption{\label{figure4}
  Electrical conductances, Seebeck coefficients and thermoelectric power factors (top, middle and bottom panels, respectively) for graphene, silicene, germanene and stanene nanoribbons with magnetic zigzag edges.}
\end{figure}

We have performed comparative calculations for graphene, silicene, germanene  and stanene using sets of parameters according to [\onlinecite{Ezawa2015}].
It should be noted that
apart from graphene these materials possess finite $t_{SO}$ and buckling parameters and have comparable lattice constants (a) and hopping integrals (t). The latter two parameters for graphene differ considerably from those for the other graphenelike materials.
This is why the results for silicene, germanene and stanene differ less from one another than from those for graphene.
This is clearly visible in Figs.~\ref{figure4}. In the IN-AP configuration and at 300K both TPFs and Seebeck coefficients always have pronounced extrema in the vicinity of the charge neutrality point.
The conductances (G) close to the CNP, in turn, are nearly zero because, as shown in Fig.\ref{FigPihM}, the systems have nonzero energy gaps. Of course at elevated temperatures the gap effects get strongly reduced and smeared.
The results for G, S and TPF in the case of graphene are similar to those reported in [\onlinecite{Karamitaheri12}] where a significant enhancement of the thermoelectric performance was achieved by introducing extended line defects (ELD). It follows from this comparison that the enhancement of the Seebeck coefficient due to edge magnetism might outperform that due to the ELD by a factor of 2, whereas the TPFs are roughly the same.  Similar conclusions result from an analogous comparison of the S-factor with that from [\onlinecite{Zheng15}] where the enhancement comes from the application of ferromagnetic leads and gating of the central region (germanene nanoribbon).

\subsection{Figure of merit}

As concerns figures of merit, following the practice of other authors, we assume that the phonon thermal conductance ($ \kappa_{ph} $) may be substantially suppressed mainly by substrates, structural imperfections, defects and isotopic inhomogeneities [\onlinecite{Seol2010,Zheng15,gunstPRB11,Wierzb15,TUDPRB15,Sevi12}].
Because this type of effects is not explicitly included in the present theory, $\kappa_{ph}$ is replaced by $\alpha \, \kappa_{ph} $ with the scaling factor ($0< \alpha \le 1$).\cite{gunstPRB11,Zberecki2013} Suppressions corresponding to $\alpha \sim 0.1$ and $\alpha \sim 0.01$ in graphene were measured in [\onlinecite{Seol2010}] and predicted theoretically in [\onlinecite{Sevi12}], respectively. The former result was due to the effect of a substrate, whereas the latter was found via a combination of geometrical structuring and isotope engineering. Moreover, a 100-fold reduction in thermal conductivity was also measured in rough silicon nanowires.\cite{Hochbaum08}

\begin{figure}[t!]
 \centering \includegraphics [width=1\columnwidth, trim=1cm 0cm 2cm 0.5cm,
clip=true] {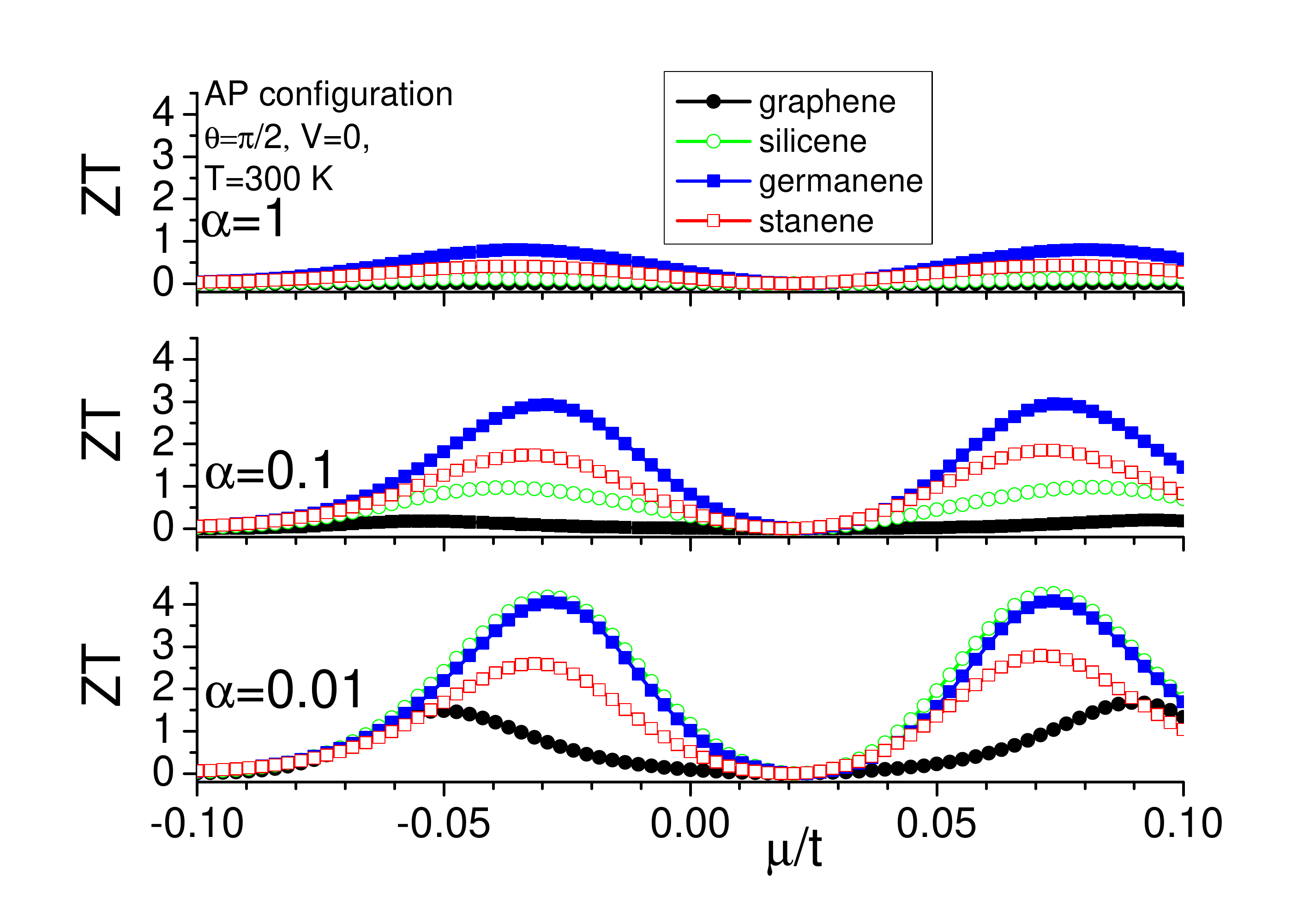}
  \caption{\label{figure5}
  Figures of merit for graphene, silicene, germanene and stanene nanoribbons with magnetic zigzag edges. Phonon contributions to the thermal conductance are scaled by the factor $\alpha$ (equal to 1, 1/10 and 1/100).}
\end{figure}

\begin{figure}[b!]
 \centering \includegraphics [width=1.0\columnwidth, trim=1cm 0.5cm 1.5cm 0cm,
clip=true] {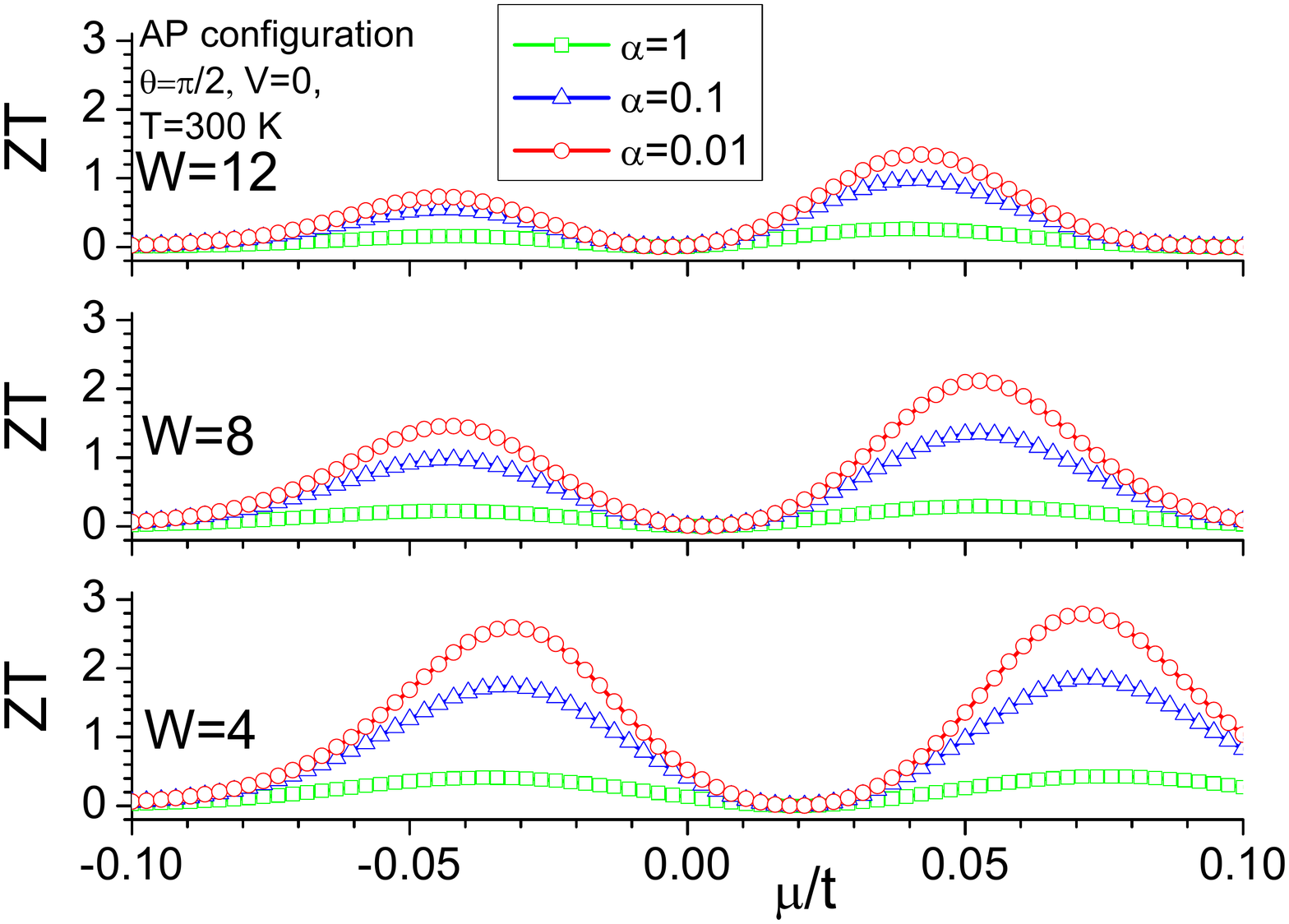}
  \caption{\label{figure6}
  Figures of merit for stanene nanoribbons with magnetic zigzag edges and the indicated scaling factors. Panels correspond to widths of 12 (top), 8 (middle) and 4 (bottom) zigzag lines, respectively.}
\end{figure}

The $\kappa_{ph}$-values for all systems studied here were taken from Ref.~[\onlinecite{Peng2016}].
Figure \ref{figure5} shows that in the case of $\alpha=1$ (no phonon suppression) only germanene has ZT nearly equal to 1, but with decreasing $\alpha$ the graphenelike nanostructures reveal quite high ZT-values.
  In particular, silicene, germanene and stanene have relatively large ZT factors attractive from the point of view of potential practical applications provided their phonon thermal conductances are reduced by 90\% or more. In fact, systems with $ZT \sim 1$ are already regarded as good thermoelectric materials, and those having $ZT \sim 3$ would be competitive with the best of the conventional energy conversion  devices. \cite{gunstPRB11}
However it should be kept in mind that experimental realization of the GLNRs of interest here is a big challenge, although at present good-quality nanoribbons of the necessary chirality can be fabricated, \cite{Tang13,Magda14,Li14} and advanced experimental techniques for electrical and thermal transport measurements are very well developed.\cite{Bauer12,Lee12,Xu14}

  The present findings show that the existence of edge magnetism improves the electrical performance of the GLNRs. On the one hand graphene which is known to be an inefficient thermoelectrical material [\onlinecite{Balendhran2015}], according to the present theory has ZT close to 1.5 (Fig.5, bottom panel). On the other hand our results for silicene and germanene are consistent with those in [\onlinecite{Yang14}] (and in [\onlinecite{Hochbaum08}] for Si nanowires)  for $\alpha=0.1$; moreover the ZT values still increase rapidly with further decreasing $\alpha$.

So far the case of ultra narrow (4 atom wide) infinitely long graphenelike nanoribbons has been considered in detail.
This case corresponds to the width of ca. 1nm (accessible experimentally [\onlinecite{Magda14, Li14}]).
 The results for ZT coefficients as a function of the nanoribbon width, presented in Fig.~\ref{figure6}, show that with increasing width the enhancement of thermoelectric characteristics becomes less and less pronounced. Noteworthily, in the case of stanene 12 zigzag lines in width (roughly 5 nm) the maximum value of ZT is still slightly above 1. From the experimental side it is known [\onlinecite{Magda14}]) that for 7 $-$ 8 nm wide graphene nanoribbons the energy gap collapses, which, according to the present theory, implies drastic worsening of thermoelectric properties.

\subsection{Gate voltage effect}

\begin{figure}[h!]
\centering \includegraphics [width=0.75\columnwidth, trim=8cm 1.4cm 8cm 1cm,
clip=true] {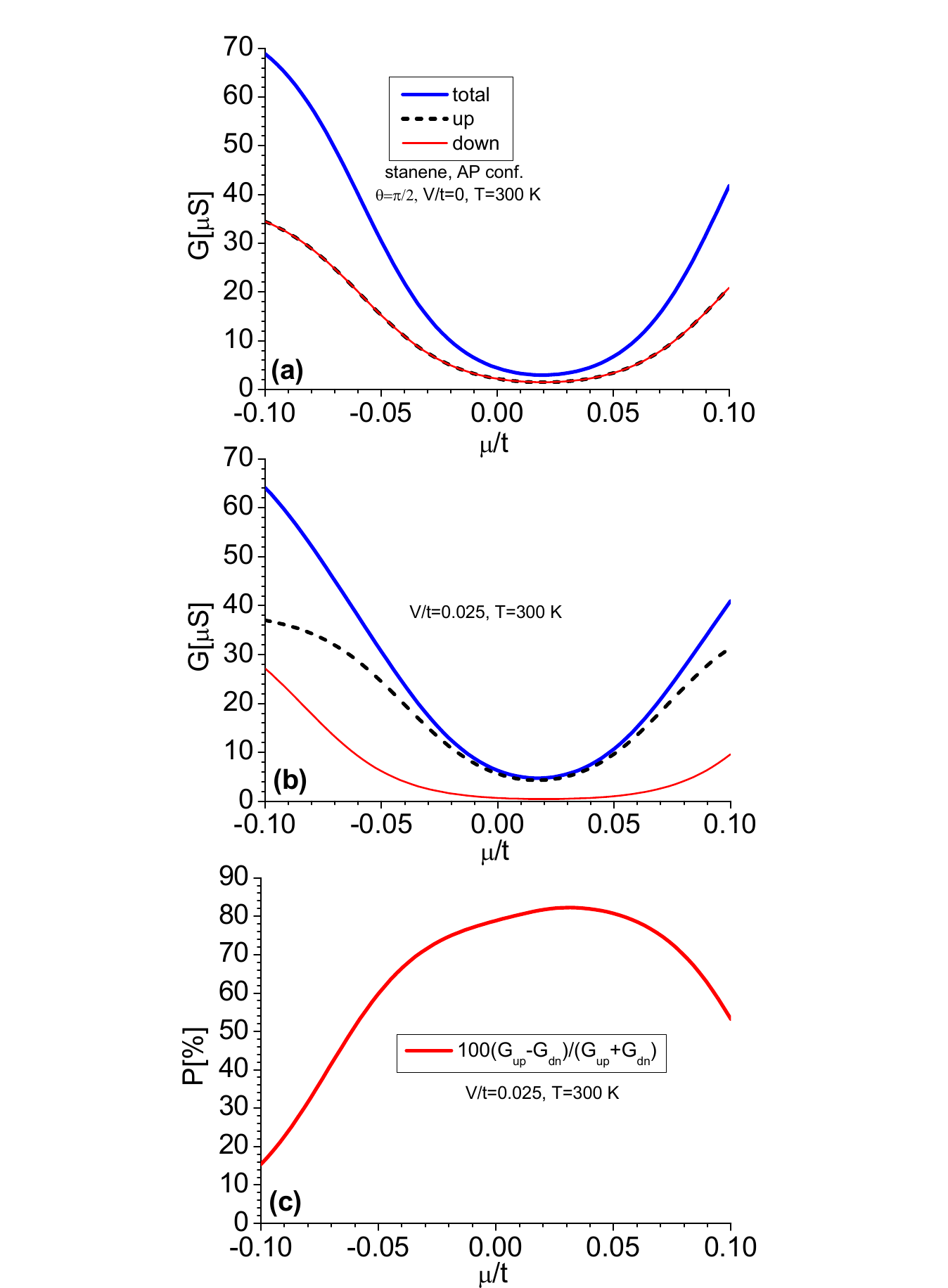}
  \caption{\label{figure7}
  Electrical conductance of stanene nanoribbons (a) without and (b) with perpendicular gate voltage. In the latter the conductance is strongly spin-polarized, the dashed-black and red curves correspond to up and down spin-contributions, whereas the blue curve represents the sum thereof. (c) Spin-polarized conductance for V/t=0.025.}
\end{figure}

\begin{figure}[h!]
 \centering \includegraphics [width=0.8\columnwidth, trim=6cm 0cm 8cm 2cm,
clip=true] {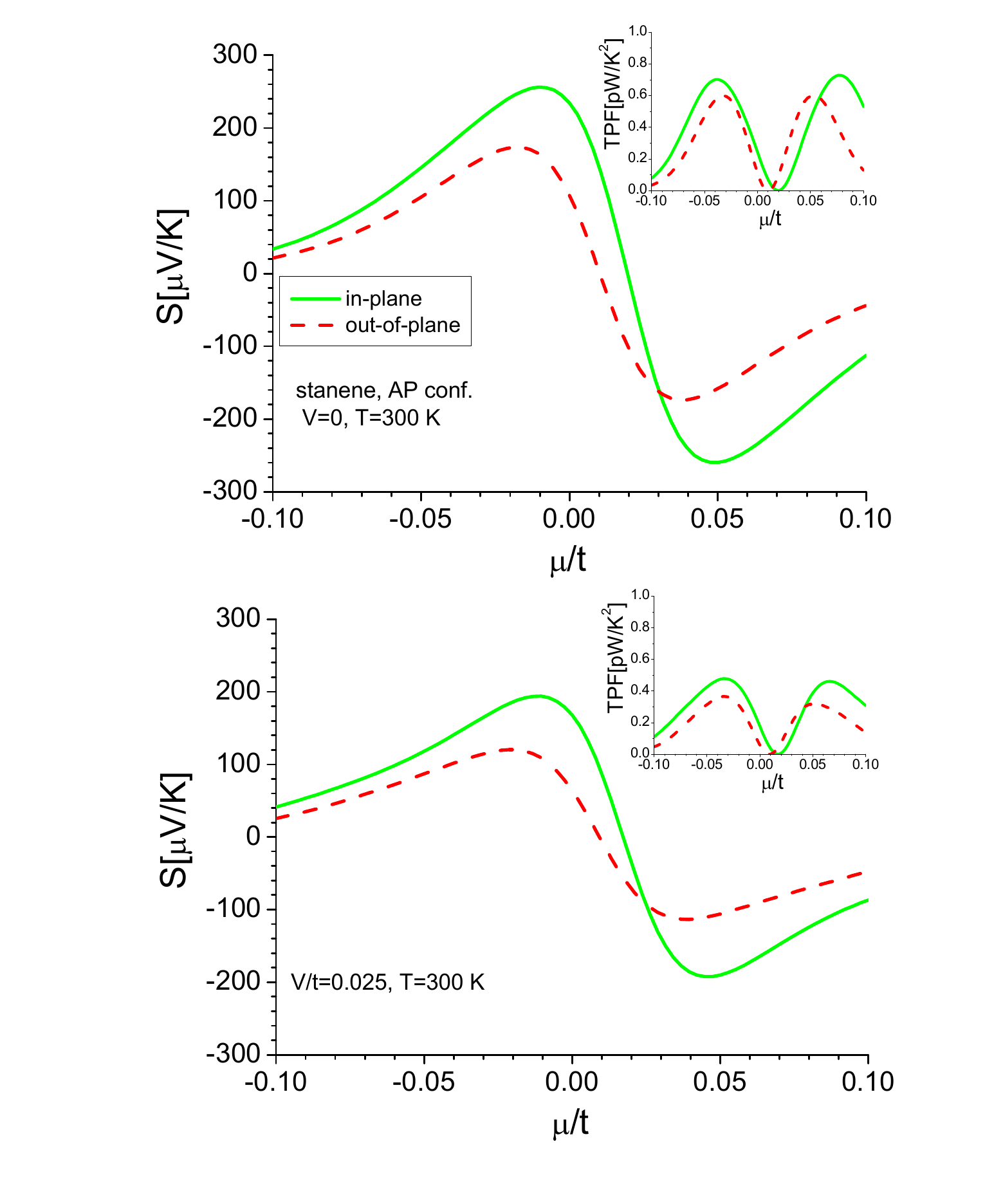}
  \caption{\label{figure8}
  Seebeck coefficient and thermoelectric power factor (inset) of stanene nanoribbon at T=300 K and indicated values of the gate voltage (V). The solid curves refer to
  the IN-AP stable phase. For comparison the corresponding curves for the OUT-AP configuration are also shown (dashed lines).}
\end{figure}

 The gate voltage effect has been intensively studied by many authors.\cite{Drummond2012,Ezawa2015,Tsai2013} If there is no edge magnetism, a small ISOI-induced energy gap gets closed for $V=V_c=3 \sqrt{3} t_{SO}$. The case of graphenelike nanoflakes with magnetic edges has been studied in Ref.~[\onlinecite{SKNanotech2016}]  and it has been found that then the critical $V_c$ is substantially increased.
Here, as an example, V/t has been set equal to 0.025 for stanene, so as to guarantee a significant spin splitting without suppressing of the edge magnetism.
 In the absence of any vertical gate voltage the electrical conductances are not spin-split in the AP arrangement (Fig.\ref{figure7}(a)), but the degeneracy is lifted at finite $V$ (panel (b)) resulting in the achievement of a relative spin-polarization of more than 80\% (panel (c)). Unfortunately, a similar trick is not possible for a graphene monolayer since then the A and B sublattice atoms are coplanar and there is no way to gate them independently. As regards the thermoelectric properties, the Seebeck coefficients (S) and the TPFs are the biggest in the case of the in-plane configuration. Moreover,  as clearly shown in Fig.\ref{figure8}, in the presence of a finite gate voltage the thermoelectric performance slightly worsens. Additionally, it should be emphasized that in the case of systems with no band gap, i.e. with parallel alignment of edge magnetizations, or non-magnetic edges, the thermoelectric performance is extremely poor.

\vspace{1cm}
\section{Summary}

In this paper we have analyzed the impact of the edge magnetism on the electrical transport
and thermoelectric properties of selected graphenelike nanoribbons (silicene, germanene, stanene, plus graphene for comparison). It has been found that the ground state of these nanoribbons corresponds to the IN-AP magnetic configuration, i.e. the in-plane antiparallel arrangement of edge magnetic moments. Close to the charge neutrality point (CNP) the systems are small gap semiconductors at low temperatures. At room temperature both the Seebeck coefficient and the thermopower factor reveal high peaks for energies in the vicinity of the CNP. The same applies to the ZT factors whose values can exceed 3 provided that their phonon thermal conductance is appropriately reduced. As concerns the perpendicular gate voltage, its effect is quite promising for possible spintronic applications connected with spin-polarized current  (cf. [\onlinecite{Tsai2013}]).

\begin{acknowledgments}
This project was supported by the Polish National Science Centre
from funds awarded through the decision No. DEC-2013/10/M/ST3/00488.
\end{acknowledgments}

\end{document}